\documentclass[a4paper]{article}
\usepackage{CJK}
\usepackage{amsmath,amsthm}
\usepackage{amsmath}
\usepackage{amssymb}
\usepackage{amsfonts}
\usepackage{eufrak}
\usepackage{eucal}
\usepackage{fancyhdr}
\usepackage{graphicx}
\usepackage{float}
\usepackage{mathrsfs}
\usepackage{graphicx,fancyhdr}
\usepackage{graphicx} 
\usepackage{CJKnumb}
\usepackage{titlesec}
\usepackage{amsthm}
\usepackage{cite}
\usepackage[utf8]{inputenc}
\usepackage[colorlinks,urlcolor=blue,citecolor=blue,linkcolor=blue]{hyperref}
\usepackage{chngcntr}
\usepackage{makecell}
\usepackage{tabularx}

\counterwithout{figure}{section}

\usepackage{flushend} 
\usepackage{ulem}    
\usepackage[numbers, sort]{natbib} 

\UseRawInputEncoding
\begin{document}
\setcounter{page}{1}
\title{AdS/CFT and the Double-Copy: A Unified Perspective Across Various Phenomenological Models}
\author{Jia-rui Guo \\ NanKai University, Tianjin 300071, China
\thanks{Email address: 2010213@mail.nankai.edu.cn}
}
\maketitle

\begin{abstract}
This study utilizes theoretical tools including the double copy relation and AdS/CFT correspondence. Applying analogies, we establish a system to discuss the associations and equivalence relations among QCD, QED, and various (mixed) gauge field theory based quantum many-body effective models. By employing mathematical techniques on the background spacetime, these relations are extended to extreme regimes while ensuring model independence.

\end{abstract}

\section{Introduction} \label{sec1}

In prior investigations, discussions of quantum chromodynamics (QCD) matter (from IR to UV) often bypassed QCD interact contribution, as direct computation from physical principles is extremely difficult, and the behavior of physical systems is influenced by various external field corrections. Certain calculations or approximate solutions under specific conditions have yielded phenomenological results \cite{1604.08082}, starting from the hadronic level, to skyrmions \cite{nov2015}, nuclear pasta \cite{1504.03964} \cite{1602.03215} \cite{2201.12598}, effective models of various quark matters(MIT bag, NJL, Lattice, Chiral Perturbation, QCD Sum Rules, etc.\cite{6}\cite{7}\cite{8}\cite{9}\cite{a}), color superconduct  \cite{9804403}, light-front holographic QCD \cite{c} \cite{Brodsky2010}, the Cornell potential \cite{Cornell}, various perturbation theory, etc. However, theoretical predictions often appear discontinuous due to being model dependent. These issues also reflect common difficulties across various quantum many-body theories.

Based on the double copy \cite{1004.0693} \cite{Bern2010} \cite{Lin2022}, AdS/CFT correspondence \cite{ac1} \cite{ac2}, and related theories, this work attempts to directly address the physical origins from more fundamental principles and universal correspondences, constructing a phenomenological mechanism that spans different interactions, effective models, and extreme spacetime backgrounds. This work allows analytical perspectives used in one context to be transferred to other models, enabling us to provide more consistent, accurate and universal conclusions, while still avoiding the difficulty of direct computation, and support physical reasoning based on justified analogies.

In part II, we harness the principles of double copy theory to yield gravity, for the perspectives of scattering amplitudes and position-space observables, respectively; and extending the applicability of the double copy relationship. In part III, we employ AdS/CFT correspondence to establish an intricate relationship, construct an effective gauge-interact-like field in AdS bulk from boundary-to-bulk propagator, and discuss the emergent duality relations therein. Lastly, part IV introduces some novel multiplication operators to generalize the relationships, and utilizing mathematical construct enables us to articulate a unified expression accommodating multiple gauge fields.

\section{General Formula for The Double Copy of various YM fields and CK Duality} \label{sec2}

\subsection{Introduction of Light front holographic QCD}

The inception of double copy theory is rooted in the correspondence between QCD and gravity. But in practice, there is a notable challenge in this domain arises when QCD coupling intensifies to such an extent that it escapes the bounds of perturbative theories, since the running coupling $a_{s}$ is limited to the perturbative or weak domain, use $g_0=g(Q=\mu)$ and define QCD scale parameter $\Lambda$ as $\Lambda^2=\mu^2 e^{\frac{-4 \pi}{\beta_0 g_0^2}}$ in $\beta_0$ order, and the running coupling $a_s = \frac{g_s^2}{4 \pi} = \frac{1}{4 \pi} \frac{g_0^2}{1+\frac{1}{\left(4 \pi \right)^2} \beta_0 g_0^2 ln \frac{Q^2}{\mu^2}}\approx \frac{4 \pi}{\beta_0 ln \frac{Q^2}{\Lambda^2}}$.

Light-front (LF) holographic QCD use holographic correspondence which is proven effective in phenomenology, and use light-front coordinate for light-front quantization, it gives an effective characterization of non-perturbative QCD \cite{Brodsky2010}, the QCD Lagrangian effective expression

\begin{equation}
\mathcal{L}_{\mathrm{QCD}}=-\frac{1}{4} F_{\mu \nu}^{a} F^{a \mu \nu}+\sum_{f} \bar{\psi}_{f}\left(i \gamma^{\mu} D_{\mu}-m_{f}\right) \psi_{f} \label{e1}
\end{equation}

Here as in LF holographic QCD the gluon field strength tensor $F_{\mu \nu}^{\alpha}$ and quark field $\psi_{f}$ are decomposed into light-front components

\begin{equation}
F_{+-}^{a}=\partial_{+} A_{-}^{a}-\partial_{-} A_{+}^{a}+g f^{a b c} A_{+}^{b} A_{-}^{c}, F_{\perp \pm}^{a}=\partial_{\perp} A_{ \pm}^{a}-\partial_{ \pm} A_{\perp}^{a}+g f^{a b c} A_{\perp}^{b} A_{\pm}^{c} \label{e2}
\end{equation}

\begin{equation}
{D}_{\mu}=\partial_{\mu}-i g A_{\mu}^{a} t^{a} \quad \psi_{f-}=\frac{1}{i \partial_{-}}\left(i \gamma^{\perp} D_{\perp}+m_{f}\right) \psi_{f+} \label{e3}
\end{equation}

\begin{equation}
g \approx \left[\frac{8 \pi^2}{\beta_0 \ln \left(\frac{Q}{\Lambda}\right)}\right]^{\frac{1}{2}}\text {pertubative, } \approx e^{-\kappa^2 z^2} \left[4 \pi a_s(0)\right]^{\frac{1}{2}} \text {LF holographic} \label{e4}
\end{equation}

the beta function in LF holographic QCD

\begin{equation}
\beta\left(a_{s}\right)=\frac{d a_{s}}{d \ln \mu}=\frac{d \left(a_s(0) e^{-2\kappa^2 z^2}\right)}{d \ln \mu}  =-\beta_{0} \frac{a_{s}^{2}}{2 \pi}-\beta_{1} \frac{a_{s}^{3}}{(2 \pi)^{2}}-\beta_{2} \frac{a_{s}^{4}}{(2 \pi)^{3}}-\cdots \label{e5}
\end{equation}

According to above, generating function

\begin{equation}
Z[J, \eta, \bar{\eta}]=\int \mathcal{D} A_{\mu} \mathcal{D} \psi \mathcal{D} \bar{\psi} \exp \left(i S_{\mathrm{QCD}}\left[A_{\mu}, \psi, \bar{\psi}\right]+i \int d^{4} x\left(J_{\mu}^{a} A^{a \mu}+\bar{\eta} \psi+\bar{\psi} \eta\right)\right)\label{e6}
\end{equation}

where $S_{\mathrm{QCD}}=\int d^4 x \mathcal{L}_{\mathrm{QCD}}$, $J_{\mu}^{a}$ is an external source for gluon field $A_{\mu}^{a}$, $\eta$ and $\bar{\eta}$ are respectively external sources for quark and antiquark fields. Then we obtain n-point green function

\begin{equation}
\langle 0| T\left\{\phi_{1}\left(x_{1}\right) \phi_{2}\left(x_{2}\right) \cdots  \phi_{n}\left(x_{n}\right) \right\}|0\rangle=\left.\frac{\delta^{n} Z[J, \eta, \bar{\eta}]}{\delta \mathcal{J}_{1}\left(x_{1}\right) \delta \mathcal{J}_{2}\left(x_{2}\right) \delta \cdots \mathcal{J}_{n}\left(x_{n}\right) }\right|_{\mathcal{J}=0}
\label{e7}
\end{equation}

Here $\phi_{n}\left(x_{n}\right)$ could be quark field $\psi,\bar{\psi}$ or gluon field $A_{\mu}^a$.

However, the running coupling and loops still cause difficulty in calculating various observables, especially in light quark interact or when the interact is strong. Also, IR and UV region have different expressions and are not unified. The following sections will try to help solving these problems.

\subsection{The Double Copy from Scattering Amplitudes}

In prior research, the L-loop m-point scattering amplitude in D-dimensional QCD and gravity has been described as a composite of propagators and vertices, also

\begin{equation}
\mathcal{M}_{m}^{(L)}=\left.\mathcal{A}_{m}^{(L)}\right|_{\substack{c_{i} \rightarrow \tilde{n}_{i} \\ g \rightarrow \kappa / 2}}=i^{L-1}\left(\frac{\kappa}{2}\right)^{m-2+2 L} \sum_{i} \int \frac{d^{L D} \ell}{(2 \pi)^{L D}} \frac{1}{S_{i}} \frac{n_{i} \tilde{n}_{i}}{D_{i}} \label{e8}
\end{equation}

There are two expressions of double copy, replacement $YM |_{\text{Gauge} \rightarrow \text{GR}} \sim GR$ and product ${YM}^2 \sim GR$. However, the replacement is only formal similarities between gauge and gravity by replacing factors, while the product simply cancels gauge factors in the second gauge amplitude via the kernel. The physical nature of these two forms seems to be different, but actually it is the same.

If the double copy is really based on physical nature of gauge and gravitational fields, it should be valid for all-loop. However, it is still difficult to prove and calculate all-loop double copy scattering amplitude, although there has been some research about this \cite{2405.09608}. The difficulties are caused by complicated structure of loop integrands and divergence of IR and UV.

But as we know, multi-loop amplitude can be reduced to lower loop level or tree-level via recursion relations and unitary-based methods. Thus, before further discussion, we should list double copy relations for cuttings in unitary-based methods, loop integrands and integrals. Feynman integrals $F_{T}^{(L),[D]}$ and integrands $I_{n}^{(L),[D]}$ are defined as following

\begin{equation}
\begin{array}{l}
A_{n}^{(L),[D]}(1, \ldots, n)=\left(g_{\mathrm{YM}} \mu_{\mathrm{R}}^{(4-D) / 2}\right)^{n-2}\left(\frac{\mathrm{i} \alpha_{\mathrm{YM}} \mu_{\mathrm{R}}^{(4-D)}}{(4 \pi)^{(D-2) / 2}}\right)^{L} \sum_{T} c_{T}^{[D]}(1, \cdots, n) F_{T}^{(L),[D]}\left(p_{1}, \cdots, p_{n-1}\right)\\
A_{n}^{(L),[D]}(1, \ldots, n)=\int \prod_{l=1}^{L} \frac{\mathrm{~d}^{D} k_{l}}{\mathrm{i}^{D / 2}} I_{n}^{(L),[D]}(k, 1, \ldots, n)
\end{array}
\label{e9}
\end{equation}

The method of maximal cuts \cite{0705.1864} \cite{1708.06807} constructs multi-loop integrands from generalized unitarity cuts.

\begin{equation}
\begin{array}{l}\mathcal{C}^{\mathrm{N}^{k} \mathrm{MC}}=\sum_{\text {states }} \mathcal{A}_{m(1)}^{\text {tree }} \cdots \mathcal{A}_{m(p)}^{\text {tree }} \\ \mathcal{C}_{\mathrm{YM}}=\sum_{\text {states }} \prod_{j=1}^{p} \sum_{\rho^{(j)} \in \mathcal{S}_{m(j)-2}} c\left(\rho^{(j)}\right) A_{m(j)}^{\text {tree }}\left(\rho^{(j)}\right) \\ \mathcal{C}_{\mathrm{GR}}=i^{p} \sum_{\text {states }} \prod_{j=1}^{p} \sum_{\rho^{(j)}, \tau^{(j)} \in \mathcal{S}_{m(j)-3}} K\left(\rho^{(j)} \mid \tau^{(j)}\right) A_{m(j)}^{\text {tree }}\left(\rho^{(j)}\right) \tilde{A}_{m(j)}^{\text {tree }}\left(\tau^{(j)}\right)\end{array} \label{e10}
\end{equation}

And $\mathcal{N}_{\text {GR}} \sim \mathcal{N}_{\mathrm{YM}}^{(1)}  \mathcal{N}_{\mathrm{YM}}^{(2)} $, for Feynman integrals computed on unitarity cuts,

\begin{equation}
F_{n}^{(1),[D]}[N]=\int_{k} \frac{N}{\prod_{a=1}^{n} D_a }\text{,and }F_{GR}^{(L)}=\int\left(\prod_{i=1}^{L} \frac{d^{d} \ell_{i}}{(2 \pi)^{d}}\right) \frac{\mathcal{N}_{\text {GR }}\left(\ell_{i}, p_{a}, \epsilon_{a}\right)}{\prod_{\alpha} D_{\alpha}\left(\ell_{i}, p_{a}\right)} \label{e11}
\end{equation}

Integrand

\begin{equation}
\mathcal{I}^{(k)} =\sum_{\ell=0}^{k-1} \sum_{\mathcal{S}_{m}} \sum_{i_{\ell}} \frac{1}{S_{i \ell}^{(\ell)}} \frac{N_{i_{\ell}}^{(\ell)}}{D_{i \ell}^{(\ell)}}\text{, similar } \mathcal{I}_{GR} \sim \sum_{\text {diagrams }} \frac{\mathcal{N}_{\mathrm{YM}}^{(1)}  \mathcal{N}_{\mathrm{YM}}^{(2)}}{\prod_{\text {propagators }} D_{\alpha}} 
\label{e12}
\end{equation}

According to previous researches (\cite{1011.1485} \cite{1107.1499} \cite{1107.1504}, etc.), similar relations can be extended to not only in Minkowski space, but also in AdS space and for AdS/CFT. Previous researches \cite{1011.1485} \cite{1107.1499} \cite{1107.1504} proved higher-loop correlators can be reduced to necklace diagram by recursion relations, and there is

\begin{equation}
\left\langle\mathcal{O}_{1} \cdots \mathcal{O}_{n}\right\rangle=\int[\mathrm{d} \delta] \mathcal{M} \prod_{i<j} \frac{\Gamma\left[\delta_{i j}\right]}{P_{i j}^{\delta_{i j}}}\text{, }\mathcal{M}[\{\Delta\},\{\underline{\Delta}\}, h]=\int \mathcal{N} M[\{\Delta\},\{c\}, h]
\label{e13}
\end{equation}

$M_{\text {new }}[s]=c \int[\mathrm{d} \Xi, \mathrm{d} \xi] M_{\text {old }}[\Xi] K[\Xi, \xi ; s]$, c is a constant. Also \cite{2106.07651} proved this relation in AdS space.

Actual matter includes not only gauge-interact terms but also mass and kinematic terms. The CK-duality can be an operator effectively merge mass and kinematic terms with gauge-interact terms. As an example, former researches using Form factors $\mathcal{F}_{\mathcal{O}, n}=\int d^{D} x e^{-i q \cdot x}\langle 12 \ldots n|\mathcal{O}(x)| 0\rangle$ \cite{Lin2022} as operators that give Fourier-transform to momentum space, thus facilitating the comparison of kinematic and color factors

\begin{equation}
\begin{aligned}
\mathcal{M}_{m}^{tree} = -i \left( \frac{\kappa}{2} \right)^2 \mathcal{G}_n =& -i \left( \frac{\kappa}{2} \right)^2 \sum_{\alpha, \beta \in S_{n-2}} \mathcal{F}_{n}[\alpha] \mathbf{S}_{n}^{\mathcal{F}}[\alpha \mid \beta] \mathcal{F}_{n}[\beta]\\
=& -i \left( \frac{\kappa}{2} \right)^2 \sum_{\alpha, \beta \in S_{n-2}} N_{n}[\alpha] \Theta_{n}^{\mathcal{F}}[\alpha \mid \beta] N_{n}[\beta]
\end{aligned} \label{e14}
\end{equation}

This can be generalized into n-point loop levels\cite{1211.7028}.

\subsection{The Double Copy of Classical Observables}

To help understand and verify the double copy relations, we need a bridge between the abstract description via amplitudes and directly classical observables. Previous researches have given the volume form of double copy \cite{1408.4434} \cite{1602.08267} as

\begin{equation}
A_{\mu}(x)(L) \otimes A_{\nu}(x)(R)  \text{, } f " \otimes " g:=f \star \Phi \star g \text{, }[f \star g](x)=\int d^{D} y\langle f(y), g(x-y)\rangle \label{e15}
\end{equation}

\begin{equation}
\mathcal{L}_{\mathrm{GR}}[A]=\frac{\alpha}{2 \lambda(\lambda-3 \alpha)}\left(\operatorname{Tr}\left(\sqrt{F^{i} \wedge F^{j}}\right)\right)^{2}+\frac{1}{2 \lambda} F^{i} \wedge F^{i} \label{e16}
\end{equation}

However, to investigate its physical nature, we should change the expression of the formula even further.

Start with the simplest case, without AdS/CFT, expand gauge interact expression, rank all the terms by $x^n$ level in position space, expand the convolution and take $f(x)=x^a$ and $g(x)=x^b$, we have

\begin{equation}
(f * g)(x)=\frac{\Gamma(a+1) \Gamma(b+1)}{\Gamma(a+b+2)} \cdot x^{a+b+1}=\frac{\Gamma(a+1) \Gamma(b+1)}{\Gamma(a+b+2)} \cdot f(x)g(x)x, \quad t \in[0, x] \label{e17}
\end{equation}

Especially, $(f * g)(x)=-mn \pi^2 \frac{1}{x}$ when $f(x)=\frac{m}{x}$ and $g(x)=\frac{n}{x}$, which satisfy the coulomb linear approximation.

When consider running coupling effects, revisit light-front holographic QCD, we notice the running coupling can rewrite as $a_s(0) e^{-\kappa^2 z^2}$.

Thus, if consider the replacement $|_{\substack{c_{i} \rightarrow \tilde{n}_{i} \\ g \rightarrow \kappa / 2}}$ as an operator $\mathcal{O}_{dc}$, it could have the form of $\mathcal{O}_{dc} \sim s A_{\nu} x$, $\phi$ is the second gauge field.

In conventional frameworks, $h_{\mu \nu}$ is employed to depict the asymptotic weak graviton field \cite{Bern2019} $g^{\mu \nu}=\eta^{\mu \nu} + \kappa h^{\mu \nu}$. However, the actual gravity is non-linear, while double copy can construct non-linear gravity\cite{1807.09859}. Write gravitational coupling

\begin{equation}
\sqrt{-g}=1+\kappa h_\mu^\mu(x)-\frac{\kappa^2}{2}\left(h^{\mu \nu}(x) h_{\mu \nu}(x)-\left(h_\mu^\mu(x)\right)^2\right)+\mathcal{O}\left(\kappa^3\right), g=|det g_{\mu \nu}| \label{e18}
\end{equation}

To delve deeper into this complexity, our focus in this section is on representing gravity as a double copy of gauge fields, either in Minkowski or AdS space. Modification in this work is to give non-linear corrections, and prior investigations have confirmed the double copy relationship for scattering amplitudes in AdS spacetime \cite{2106.07651}.

Importantly, we examine gravity coupled to a tensor field constructed by double copy rather than a scalar field, in order to get closer to its physical nature, we should construct entire energy-momentum tensor from double copy.

\begin{equation}
T_{\mu\nu} = \frac{\partial \mathcal{L}}{\partial (\partial^\mu \phi)} \partial_\nu \phi - g_{\mu\nu} \mathcal{L}\text{, } \mathcal{L}=k \left(\operatorname{Tr}\left(\sqrt{F^{i} \wedge F^{j}}\right)\right)^{2}+\frac{1}{2 \lambda} F^{i} \wedge F^{i} \label{e19}
\end{equation}

Review previous work, the double copy of currents has been proved, which has given a direct support for this. The double copy of currents is given by\cite{2109.06392}

\begin{equation}
\mathcal{J}_{\mu \nu}^{\mathcal{P}}=\sum_{\alpha, \beta \in S_{\mathcal{P}^{*}}} J_{\mu}^{1 \alpha} S[\alpha \mid \beta]_{1} J_{\nu}^{1 \beta} \text{, quark current } J^{\mu}=e_q \tilde{q} \gamma^{\mu} q \text{ and } J^{\mu} =\lim_{r\rightarrow \infty}\frac{\delta S}{\delta A_{\mu}} \label{e20}
\end{equation}

Within the context of $AdS_5$, form factors are also applicable to hadronic QCD fields$A^M(x, z)$. Use $\Phi(z)$ for hadron field, $\left(4 \pi g_{LF}^2 N_c \right)^{\frac{3}{4}}l_s^3$, $\frac{1}{l_s^2}$ stands for string scale ,$z=\frac{1}{\Lambda_{QCD}}$ and M for indices $x^M=(x^{\mu} ,z)$.

\begin{equation}
\begin{aligned}
\mathcal{F} (q^2)= & \left(4 \pi g_{QCD}^2 N_c \right)^{\frac{3}{4}}l_s^3 \int_0^{\frac{1}{\Lambda_{QCD}}} \frac{dz}{z^3} \frac{e^{-iqx}}{\epsilon_{\mu}(q)} A_{\mu}(x^{\mu}, z) \Phi^2(z)\\
\mathcal{M}_{m}^{(L)} \sim & \left(4 \pi g_{QCD}^2 N_c \right)^{\frac{3}{2}}l_s^6 \left[\int dz \sqrt{g} |det g_{\mu \nu}| \frac{e^{-i q_{\alpha} \cdot x}}{\epsilon_{\mu}(q)} A_{\alpha \mu}(x^{\mu}, z) \Phi_{\alpha}^2(z)\right] \left(\Theta_n^{\mathcal{F}}\right)^{-2} \\
& \left[\int dz \sqrt{g} |det g_{\mu \nu}| \frac{e^{-i q_{\beta} \cdot x}}{\epsilon_{\mu}(q)} A_{\beta \mu}(x^{\mu}, z) \Phi_{\beta}^2(z) \right] \\
\end{aligned} \label{e21}
\end{equation}

Noticing $\sqrt{g} \sim \frac{R}{z}$ in AdS spacetime, amplitude expressions can be reformulated as above.

Therefore, we obtained a possible form of non-linear gravity from loop-level double copy and with its source a complete double copied tensor field($T_{\mu \nu}$), which comes from two gauge fields with current $J_{\mu}$. For its applicable condition to be physical, see the next section.

\subsection{Double Copy as Product and Modification}

Construct a special kind of double copy: The gravity $g_{\mu \nu}(x)$ comes from gravity coupling with energy-momentum tensor $T_{\mu \nu} (x)$ of matter field instead of pure gravity, and $T_{\mu \nu} (x)$ comes from gauge field $A_{\mu} (x)$ modified by another gauge field (which is the partner in double copy), two "x" above is the same variable for position, and therefore there should be a correspondence between the correlation function of gauge field $A_{\mu} (x)$ and that of gravity field $g_{\mu \nu}(x)$. In this case we define relations as

For current,

\begin{equation}
T^{\mu\nu} = \frac{\partial \mathcal{L}}{\partial (\partial_\mu \phi_i)} \partial^\nu \phi_i - g^{\mu\nu} \mathcal{L}, \mathcal{L}_{\text{int}} = j^{a\mu} A_\mu^a(x), A^{a\mu}(x) = \int d^4y \, G^{\mu\nu}(x - y) J^{a}_{\nu}(y) \label{e22}
\end{equation}

Where the color current $j^{a\mu} =\bar{q} \gamma^\mu T^a q + f^{abc} F^{b\mu\nu} A_\nu^c$, and for the field,

\begin{equation}
g_{\mu \nu} (x)= A_{\mu}(x) \star \Phi_A(x,y) \star \tilde{A}_{\nu}(x-y) =\mathcal{O} A_{\mu}(x) \star \Phi_A(x,y) \star (\mathcal{O}^*)^{-1}  \tilde{A}_{\nu}(x-y) \label{e23}
\end{equation}

Where $\langle,\rangle$ is Killing form, $[f \star g](x)=\int d^{D} y\langle f(y), g(x-y)\rangle$, $\mathcal{O}^*$ is adjoint operator $\langle \mathcal{O} a, b \rangle = \langle a, \mathcal{O}^* b \rangle, \forall a, b \in \mathfrak{g}$. For simplification, we use light front holographic QFT to encode loop contribution (thus no direct loop-contribution in the expression)in LF-running coupling effect, utilizing operator $\mathcal{O} \sim \frac{\beta(Q) A_{\mu, \text{original}}}{\sum_{L=1}^L A_{\mu, \text{L}}}$. Revisit equation(5) in section II.1, the operator separately coupling with various loop levels contribution of gauge field

\begin{equation}
g_{\mu \nu} (x)=  A_{\mu, L}^{\text{LF}}(x) \star \Phi_A(x,y) \star \sum_{L=0}^L \tilde{A}_{\nu, L}(x-y) (\mathcal{O}^*)^{-1}_L(\beta), A_{\mu, L}^{\text{LF}} =\sum_{L=0}^L A_{\mu, L} \mathcal{O}_L(\beta) \label{e24}
\end{equation}

Where L=0 stands for tree-level for short, $\beta(a_s) = - \frac{Q^2}{4\kappa^2} a_s $.

According to above, and revisit the double copy of currents in section II.3 and double copy relation of field $A_{\mu}$, comparing with $A^{a\mu}(x) = \int d^4y \, G^{\mu\nu}(x - y) J^{a}_{\nu}(y)$ which has the form of convolution, we give $T_{\mu \nu}^{\text{interact}} (x) \sim j_{\mu}(x) \star \Phi_j(x,y) \star \tilde{j}_{\nu}(x-y) $ for the current, $\Phi_j$ is implied to be the zeroth-copy kernel $S[\alpha \mid \beta]$.

Thus, a double copy partner $\tilde{A_{\nu}}$ with loop contribution is designed for a gauge interact $A_{\mu}$ whose loop contribution emerged into effective running coupling. Especially, when $A_{\mu} (x)$ shares the same coordinates with gravity $g_{\mu \nu} (x)$, it could be understood as $\tilde{A_{\nu}}$ provides an modification operation via double copy relations, to enable a local relation between $A_{\mu} (x)$ and $g_{\mu \nu} (x)$, as what $|_{\substack{c_{i} \rightarrow \tilde{n}_{i} \\ g \rightarrow \kappa / 2}}$ does, but directly based on their physical nature. This change could utilize the double copy "product"(in momentum space) to construct a local "modification"(in position space) for one of the two gauge fields involved in the double copy.

This is the exact reason why here we use position space view instead of momentum space. Momentum space make it easier to calculate the amplitudes, but position space is to ensure whether the relationship is physically local or not. Since their kernel, current and field satisfy the requirements for double copy, we can extend this conclusion, start from deriving correlation function from currents

\begin{equation}
G^{\mu_1 \mu_2 \cdots \mu_n}(x_1, x_2, \dots, x_n) = \langle \Omega | T\{ J^{\mu_1}(x_1) J^{\mu_2}(x_2) \cdots J^{\mu_n}(x_n) \} | \Omega \rangle \label{e25}
\end{equation}

And double copy relation

\begin{equation}
\mathcal{G}^{\mu\nu,\alpha\beta}(q) = \int d^d x \, e^{i q \cdot x} \langle \Omega | T\{ T^{\mu\nu}(x) T^{\alpha\beta}(0) \} | \Omega \rangle_{\text{grav}} \label{e26}
\end{equation}

And form factors

\begin{equation}
F^{\mu}(p, p'; S, S') = \langle p', S' | J^\mu(0) | p, S \rangle, \mathcal{F}^{\mu\nu}(p, p') = \kappa \, F_1^{(\mu}(p, p') \, F_2^{\nu)}(p, p') \label{e27}
\end{equation}

And gravitational field

\begin{equation}
g_{\mu \nu} (x)= g^{(0)}_{\mu \nu} + \sum_{l=1}^{L} G_N^l \, g^{(l)}_{\mu \nu} + O(G_N^{L+1}) \label{e28}
\end{equation}

Given the expression of the partner field $\tilde{A_{\nu}} (x)$

\begin{equation}
\tilde{A_{\nu}} (x)=\frac{1}{x} \sum_{n=0}^{\infty} \alpha^n \sum_{m=0}^{n} c_{m n} (\ln (\mu x))^m +k \langle g^2 F_{\mu\nu}^a F^{a\mu\nu} \rangle \lambda_c^2 x \label{e29}
\end{equation}

\begin{equation}
g_{\mu \nu} (x)= A_{\mu, L}^{\text{LF}}(x) \star \Phi_A(x,y) \star \sum_{L=0}^L \frac{1}{x-y} \sum_{n=0}^{\infty} \alpha^n \sum_{m=0}^{n} c_{m n} (\ln (\mu (x-y)))^m +k \langle g^2 F_{\mu\nu}^a F^{a\mu\nu} \rangle \lambda_c^2 (x-y) (\mathcal{O}^*)^{-1}_L(\beta) \label{e30}
\end{equation}

Thus the modification from the partner $\tilde{A}_{\nu}$ derived from

\begin{equation}
A_{\mu}|_{\substack{c_{i} \rightarrow \tilde{n}_{i} \\ g_{\mu \nu} (x)}}=  \sum_{L=0}^L A_{\mu, L} \mathcal{O}_L(\beta)(x) \star \Phi_A(x,y) \star \sum_{L=0}^L \tilde{A}_{\nu, L}(x-y) (\mathcal{O}^*_L)^{-1}(\beta) \label{e31}
\end{equation}

\section{Double Copy in AdS gravity and the Conformal Field} \label{sec3}

\subsection{Double Copy in AdS/CFT}

In AdS/CFT correspondence we have \cite{1407.8131}

\begin{equation}
Z_{C F T}[j]=\left\langle\exp \left(i \int d^d x j(x) \mathcal{O}(x)\right)\right\rangle \sim Z_{\text{AdS} }[J, \eta, \bar{\eta}]  \label{e32}
\end{equation}

Previous researches have discussed double copy relation for AdS and for 3-point correlators of currents \cite{1812.11129}. Take the logarithm, we have

\begin{equation}
S_{\text{bulk}}[\phi] \big|_{\text{on-shell}}=W[J] = \sum_{n=1}^\infty \frac{1}{n!} \int d^d x_1 \cdots d^d x_n \, \langle \mathcal{O}(x_1) \cdots \mathcal{O}(x_n) \rangle_{\text{conn}} \, J(x_1) \cdots J(x_n) \label{e33}
\end{equation}

Where $S_{\text{bulk}}[\phi] \big|_{\text{on-shell}} = S_{\text{bulk}}[\phi_{\text{sol}}]$, $\phi_{\text{sol}}$ is a solution to the equations of motion, a field $\phi(x,z)$ in the bulk relates to its boundary value $\phi_0(x)$ by $\phi(x, z) = \int d^d x' \, \tilde{V}\left(Q^2, z\right) \, \phi_0(x')$. W[J] is a series of connected correlation functions, as it directly corresponds to AdS action, it can also include a structure for double copy. Previous research has given double copy relation of 3-point correlators of currents \cite{2106.07651}

In the on-shell scenario, gauge fields undergo modifications via the bulk to boundary propagator $\tilde{V}\left(Q^2, z\right)$\cite{1407.8131}, which assumes a value 1 at zero momentum transfer

\begin{equation}
\tilde{V}\left(Q^2, z\right)=|\lambda| z^2 \int_0^1 \frac{d x}{(1-x)^2} x^{Q^2 / 4|\lambda|} e^{-|\lambda| z^2 x /(1-x)} \label{e34}
\end{equation}

\begin{equation}
A_{\mu}(x^{\mu}, z)= e^{i q \cdot x }\int_0^1 dx J_0 \left( zQ \sqrt{\frac{1-x}{x}} \right) \epsilon_{\mu}(q) \label{e35}
\end{equation}

The current J, denoted earlier, refers to the quark current. When extended to on-shell AdS spacetime, this modification has implications for the double copy relationship

\begin{equation}
\mathcal{J}_{MN}^{\mathcal{P}}=\sum_{\alpha, \beta \in S_{\mathcal{P}^{*}}} \tilde{V}\left(Q^2, z\right) J_{M}^{1 \alpha} S[\alpha \mid \beta]_{1} \tilde{V}\left(Q^2, z\right) J_{N}^{1 \beta} \label{e36}
\end{equation}

To account for dilaton term, we employ an alternative strategy that involves adjusting the metric through the incorporation of a warp factor

\begin{equation}
\tilde{g}_{MN}=e^{2\tilde{\varphi} (z)} g_{MN}=e^{\frac{2 }{3} \lambda z^2} g_{MN} \label{e37}
\end{equation}

Within the AdS bulk, the dilaton modulates the contributions of the kinetic and mass density terms, thus aligning their form with that of the gauge interaction potential contributions. This alignment is achieved because the gauge interaction potentials already implicitly capture a dilaton background within the running coupling constant.

\begin{equation}
S_{\text{on-shell}}= \frac{1}{2\kappa^2} \int d^dx \sqrt{-g} \left[\left( R -2 \Lambda \right)-\frac{1}{4}  e^{-\Phi(z)} F_{MN} F^{MN}+ \mathcal{L}_{\text{int}} \left( \phi, A_M, \psi \right) \right] \label{e38}
\end{equation}

While the boundary-to-bulk propagator modifies the form factors in AdS bulk, which is effectively

\begin{equation}
\mathcal{F}(0, z) = F_0 C_{\Delta}  e^{-\kappa^2 z^2 / 2}  (\kappa z)^{2\Delta}  U\left(\frac{\Delta}{2}, \Delta, \kappa^2 z^2\right) \label{e39}
\end{equation}

Where $C_{\Delta} = \frac{\Gamma(\Delta/2) \kappa^{-\Delta}}{c_{\Delta} \Gamma(\Delta)}\),\(c_{\Delta} = \frac{\Gamma(\Delta)}{\pi^{d/2} \Gamma(\Delta - d/2)}$, and U is Tricomi confluent hypergeometric function

\begin{equation}
\mathcal{F}(0, z) \approx F_0  e^{-\kappa^2 z^2 / 2} \cdot \mathcal{F}\left(\kappa^2 \frac{x^2 z^2}{x^2 + z^2}\right) \label{e40}
\end{equation}

Where $\mathcal{F}(v) = C_{\Delta}  \left(\kappa^2 (x^2 + z^2)\right)^{\Delta} U\left(\frac{\Delta}{2} + \frac{v}{4\kappa^2 (x^2 + z^2)}, \Delta, v\right)$, or long-distance approximation $R(x, z) \approx D_{\Delta}  e^{-\kappa^2 z^2 / 2}  \left(\frac{z}{x}\right)^{\Delta} e^{-\kappa |x| z / \sqrt{2}}$, $D_{\Delta}$ is a constant.

\subsection{Double copy of Scattering Amplitudes in AdS/CFT}

In this section we will compare it with action in CFT with conformal field strength. The purpose of using AdS/CFT correspondence here is to establish a quantitative relationship (as implied in part II) between the gravitational action and its counterpart in CFT, not only a formal similarity.

When examining the correspondence between the generating functional within the framework of AdS/CFT and the double copy relationship of amplitudes, valuable insights emerge.

\begin{equation}
Z_{CFT}[j]=Z_{grav}[\Phi_{z=\epsilon} \rightarrow j]  \label{e41}
\end{equation}

The double copy in AdS has been discussed in part II, while in conformal field theory, we use Mellin amplitude (Revisit eq.13) as CFT correlator, j as source, U as derivative of external source, N is normalization factor

\begin{equation}
\begin{aligned}
Z_{CFT}[j]= & \frac{1}{N} e^{U} exp \left[ -i \int d^4 x d^4 y j^{a \mu}(x) G_{\mathrm{bb}}^{\underline{\Delta}}\left[X_{1}, X_{2}\right] j^{b \nu}(y) \right] \text{, and} \\
Z_{CFT}[j] =& \left\langle\exp \left(i \int d^d x j(x) \mathcal{O}(x)\right)\right\rangle = exp W[J] = exp S_{\text{bulk}}[\phi] \big|_{\text{on-shell}}
\end{aligned}  \label{e42}
\end{equation}

$G_{\mathrm{bb}}^{\underline{\Delta}}\left[X_{1}, X_{2}\right]$ is bulk-to-bulk propagator in the AdS corresponded to this CFT. According to discussion about double copy in part II and AdS/CFT,

\begin{equation}
\text{AdS bulk: }A_{\mu}(x,z)* A_{\nu}(x,z)e^{-\phi(z)} \sim \text{correspond to } A_{\mu}^{\text{CFT}}(z, x) = \int d^d x' \, K(z, x; x') J_\mu(x') \label{e4c}
\end{equation}

The right side of equation above is a convolution of boundary current and boundary-to-bulk propagator which can viewed as an effective correlator $\mathcal{O}_{\partial-b}$. Neglecting mass as discussed in part II,

\begin{equation}
\int d^d x \left[ \mathcal{O}_{\text{CFT}} \left (A_{\mu}^{\text{CFT}}(z, x) \right) * \mathcal{O}_{\partial-b} \right]  =\int d^d x dz A_{\mu}(x,z)\star \Phi \star A_{\nu}(x,z)e^{-\phi(z)} \label{e44}
\end{equation}

Compared to previous result, it has a double-copy-like form. Further factorize $\mathcal{O}_{\partial-b}$ and rewrite as

\begin{equation}
\int d^d x \left[ \mathcal{O}_{\text{CFT}} \left (A_{\mu}^{\text{CFT}}(z, x) \right)\star \Phi_{\text{CFT}} \star A_{\partial-b,\text{eff}} \right]  =\int d^d x dz A_{\mu}(x,z)\star \Phi_{\text{AdS}} \star A_{\nu}(x,z)e^{-\phi(z)} \label{e45}
\end{equation}

$A_{\partial-b,\text{eff}}$ can have a similar form with confine potential $A(x)=\frac{R^{2}}{\alpha^{\prime}} \frac{e^{2 f(b)} U^{4}(b, 0, f(b))}{2 \pi f(b)} c x$, U is Tricomi confluent hypergeometric function(f(b) is non-trivial function of b to ensure energy and distance take real values, see \cite{2112.00021} \cite{2209.07109}

The left side of equation above stretches across AdS boundary and bulk, that two sides of equation come together in a similar form. When the trace of bulk AdS can be transformed to vanish as described in part II, two sides of this equation can exchange when describing a random point (x, z) in AdS bulk.

Further, higher loop-level contributions in AdS bulk with higher-order couplings require dilaton modifications with negative exponents, where the power k is the negative of the power of the coupling constant as it approaching to boundary in the operator, the dilaton modification for L-loop level should be $\mathcal{O}_L \sim e^{-L \Phi}$, dilaton provides a suitable carrier for the operator required in part II. Rewrite the expression

\begin{equation}
g_{\mu \nu} (x)=  \sum_{L=0}^L A_{\mu, L} \left(e^{-L \phi} \right)(\beta)(x) \star \Phi_A(x,y) \star \sum_{L=0}^L \tilde {A}_{\partial-b,\nu, L,\text{eff}}(x-y) (\left(e^{-L \phi} \right)^*)^{-1}(\beta)  \label{e46}
\end{equation}

The two operators satisfies relations in Part II, which acts as a bridge for dual-relationships in double copy and AdS/CFT, which will be explained later.

The double copy describes the scattering amplitude aspect of the duality, AdS/CFT gives the modifier for form factors and dilatonic operator separately acts on different loop-level contribution in need, and also ensures the relationship of generating function or action of both sides of the equation. Not only its way of distribution but also exact scale of value of a given gravitational field can be constructed by the specific type of double copy described above.

Start from the equivalence of form factors and scattering amplitudes, we can further explore the equivalence of correlation function.

\subsection{Correlation Functions and Equations of States}

From AdS/CFT we know

\begin{equation}
\left\langle \exp\left( \int_{\partial \text{AdS}} d^d x \, A_\mu^{(0)}(\mathbf{x}) J^\mu(\mathbf{x}) \right) \right\rangle_{\text{CFT}} = \exp\left( -S_{\text{ren}}[A_{\text{cl}}] \right) \label{e47}
\end{equation}

Former research has discussed action of CFT \cite{1805.12100}

\begin{equation}
S_{\text{ren}}[A_{\text{cl}}] = \int_{\text{AdS}} d^{d+1}x \, \sqrt{g} \left( -\frac{1}{4e^2} F_{\mu\nu} F^{\mu\nu} \right) + \int_{\partial \text{AdS}} d^d x \, \sqrt{\gamma} \left( \text{boundary terms} \right) + S_{\text{ct}}[A_{\text{cl}}] \label{e48}
\end{equation}

Where $\sqrt{g}$ and $\sqrt{\gamma}$ are the bulk and induced boundary metrics. AdS/CFT provides the relations between generating function of CFT and action in AdS, given

\begin{equation}
\begin{aligned}
\mathcal{L}_{\text{bulk}}(x, z) = & C \left( \frac{L}{z} \right)^5 e^{-\kappa^2 z^2} \mathcal{L}_{\text{CFT}}(x, \mu), \mu \sim z^{(-1)}\\
T_{MN}^{\text{bulk}}(x,z) = & e^{-\kappa z^2} \int d^d x' \, \mathcal{K}_{MN}^{\mu\nu}(z, |x-x'|) \mathcal{T}_{CFT\,\mu\nu}(x')
\label{e49}
\end{aligned}
\end{equation}

Where $K(z,x;x') \propto e^{-\kappa z^2} z^{\Delta} U(a,b, \kappa z^2) \frac{1}{(z^2 + |x-x'|^2)^{\Delta}}$, and

\begin{equation}
Z[h] = \int \mathcal{D}\phi  \exp\left( i \int d^4 x  \mathcal{L}(x) + i \int d^4 x  h_{\mu\nu}(x) \mathcal{T}^{\mu\nu}(x) \right) \label{e50}
\end{equation}

The generating functional $Z[J]$ can be used to derive all correlation functions and make theoretical predictions for various quantum states of matter; whereas the correlation functions can be expressed in a form manifestly explicit in the effective coupling $g_{eff}$ and the form factor distribution $F(x)$.

\begin{equation}
G^{(n)}(x_1, \dots, x_n)= \left( g_{\text{eff}} \right)^{n/2} \sum_{\text{pairings}} \prod_{\text{pairs } (i,j)} \left[ \int d^4z  F(z)  G^{(2)}(x_i, z) G^{(2)}(x_j, z) \right] \label{e51}
\end{equation}

Where n is even number. Within QCD-based cases, for hadronic state we have

\begin{equation}
F(r) \approx \frac{C}{r} e^{-m r} U\left(1, 0, 2m r\right), m \sim \Lambda_{\text{QCD}} \label{e52}
\end{equation}

C is normalization constant. And for nuclear pasta

\begin{equation}
F(\mathbf{r}) \approx \sum_n C_n \frac{e^{-\kappa_n |\mathbf{r} - \mathbf{R}_n|}}{|\mathbf{r} - \mathbf{R}_n|} U\left(1, 0, 2\kappa_n |\mathbf{r} - \mathbf{R}_n|\right) \label{e53}
\end{equation}

Various nuclear pasta states can be regarded as a transition phase from the hadronic state to the color-superconducting state, corresponding to an IR-to-UV crossover in energy scales, and the configurations of various nuclear pasta states can be constructed within a finite-volume box \cite{2201.12598}. When it turns to completely color-superconducting state, we have

\begin{equation}
F(\mathbf{r}) = F_0  \frac{e^{-\kappa |\mathbf{r}|}}{|\mathbf{r}|}  U\left(1,\  0,\  2\kappa |\mathbf{r}|\right) \label{e54}
\end{equation}

Here $\kappa$ stands for chromomagnetic screening mass $\kappa = \sqrt{\frac{g^2 \mu_q^2}{2\pi^2} \left(1 - \frac{\Delta^2}{\mu_q^2} K\left(\frac{\Delta}{\mu_q}\right)\right)}$ and $K(x) = \int_0^\infty \frac{\sin t}{t} e^{-x t} dt$. In the context of constructing an effective theory for quasi-excitations in QCD matter under kinematic and external field corrections, the Tricomi confluent hypergeometric function $U(a,b,\zeta)$ provides an analytical ansatz to characterize the spatial form factor of emergent quasi-structures. Specifically, it encodes the interplay between exponential screening and algebraic decay in the density distribution of topological solitons within nuclear pasta phases.

Within QED-based cases, the role of quasiparticles in connecting QED to effective quasiparticle theories can be analogized to how hadrons and various effective models (e.g., MIT bag model, NJL model, etc.) connect QCD to phenomenological theoretical results. Quasiparticle effective theories do not purely derive from electrodynamics but emerge from the combined effects of electrodynamic factors and kinematic factors. Similarly, all effective models incorporate kinematic factors (including mass terms). However, under color-kinematic duality, we can effectively describe kinematic factors as a external field correction. By describing the electromagnetic form factors of quasiparticles, the quantum many-body complexities that prevent direct calculation via QED can be circumvented. Likewise, QCD effective models encapsulate complexities arising from loop diagrams, gluon self-interactions, color confinement, and other effects. And in many cases, the electromagnetic form factors for quasiparticles in quantum many-body theory, can be rewrite as the form that includes a constant, an exponential decay, a modulating function $\rho(r)$, and Tricomi confluent hypergeometric function $U(a,b,\zeta)$

\begin{equation}
F(\mathbf{r}) = F_{0} e^{-\kappa r} \rho(r) U(a,b,\zeta) \label{e55}
\end{equation}

Where $\kappa$ characterizing the screening length scale.

According to above, we can construct a correspondence between various states of matter (with stronger or weaker coupling), the relationship between their correlation function ensured by AdS/CFT via Z[J] and the difference described by $U(a,b,\zeta)$ and $\Lambda_{\text{eff}}$ can be given by specific boundary-to-bulk propagators and dilaton in AdS/CFT which actually acts as the partner in the specific double copy discussed in part II.

\subsection{Equivalence of Field Strengths and Introduction of Transformative Modifiers in $AdS_5$ and CFT Domains}

In the context of string theory, AdS/CFT embodies S-duality, while color-kinematics duality embodies T-duality. However, by cross-referencing and complementing the more universal double copy and AdS/CFT correspondences—which are less strictly dependent on the full requirements of string theory—and considering their physical implications, one can preliminarily construct a degenerate version that more closely approximates a general quantum field theory (with some additional string-theoretic ingredients). This version exhibits equivalence under transformations where particles and quasiparticles exchange roles in processes with different coupling strengths (strong-weak transformations, or from IR to UV), and under scale exchanges between kinematic factors and charge factors. These equivalences are themselves founded upon a change in the shape of the physical background concurrently with a transformation of physical parameters according to the rules of a specific group. This degenerate framework is more phenomenologically oriented, thereby facilitating easier verification and modification. And the purpose of this subsection is to derive the transformation according to above.

Generally speaking, it is easier to derive the energy-momentum tensor, action, generating functional, and correlation functions from the Lagrangian, whereas the reverse is more difficult. Therefore, we will proceed from the Lagrangian here. In AdS/CFT \cite{1407.8131} \cite{1301.1651} when AdS metric is a warped metric,

\begin{equation}
d s^{2}=\tilde{g}_{M N} d x^{M} d x^{N}=\frac{R^{2}}{z^{2}} e^{2 \tilde{\varphi}(z)}\left(\eta_{\mu \nu} d x^{\mu} d x^{\nu}-d z^{2}\right) \label{e56}
\end{equation}

According to double copy expression, consider the source of gravity and gauge interact,

\begin{equation}
\mathcal{L} = \frac{\sqrt{-g} \, e^{-\phi(z)}}{\kappa^2 (d-1)} \left( 2\Lambda - \kappa^2 T^\mu_\mu \right) , T^\mu_\mu = T_0 +c J^{a\mu} A^a_\mu \label{e57}
\end{equation}

c is coefficient, $T_0$ is trace anomaly (given by double copy partner in this case, according to discussion above) and mass contribution, where $T_0 = \frac{\beta(g)}{2g} F^{a\mu\nu} F^a_{\mu\nu} + i \sum_i \bar{\psi}_i \gamma^\mu \partial_\mu \psi_i $, and $T_0$ can be combined into similar terms of $c J^{a\mu} A^a_\mu$ with the help of CK duality(which will be introduced later). And for the gravity field $g^{MN} =A_{M, L}^{\text{LF}} \star \Phi_A(x,y) \star \sum_{L=0}^L \tilde{A}_{N, L}(x-y) (\mathcal{O}^*)^{-1}_L(\beta)$ where $A_{M, L}^{\text{LF}} =\sum_{L=0}^L A_{M, L} \mathcal{O}_L(\beta)$

\begin{equation}
\mathcal{L} = \frac{1}{2\kappa^2} \sqrt{-g}  e^{-2\Phi}  \left[ R + 4  g^{MN}(z, x)  \partial_M \Phi  \partial_N \Phi + \frac{2d(d-1)}{L^2} - V(\Phi) \right] \label{e58}
\end{equation}

According to boundary-to-bulk propagator effective expression

\begin{equation}
\mathcal{L} = \sqrt{-g} \, e^{-\varphi(z)} \left[ \mathcal{L}_0(\Phi) + g \, F(x, z) \, J_M^a[\Phi] \, A^{a M} \right] \label{e59}
\end{equation}

Where

\begin{equation}
\mathcal{L}_{\text{int}} = j^{aM}(x, z) A_M^a(x, z),\text{ and } A^{aM}(x,z) = \int d^4y \, G^{MN}(x - y) J^{a}_{N}(y,z) \label{e60}
\end{equation}

Here $F(x, z)$ is form factor modified by boundary-to-bulk propagators, which can often write as $F(x, z) \sim F_0 (x, z) \frac{C}{r_{\text{eff}}} e^{-\epsilon_{\text{eff}} r} U(a,b,z) $, U(a,b,z) is Tricomi confluent hypergeometric function. This can often describe effective model structures (such as effective fields or quasi particles) located in and interact with their physical background (such as background field)

More over, boundary-to-bulk propagators can generate a field in bulk AdS from a boundary source J, thus describing a correspond field in AdS bulk, which can be described as a (potential) field $\phi$

\begin{equation}
\phi(z, \vec{x})=J_{0} \cdot \Gamma\left(\frac{\Delta}{2}\right) \frac{z^{\Delta} e^{\kappa z^{2} / 2}}{\left(z^{2}+|\vec{x}|^{2}\right)^{\Delta / 2}} \cdot U\left(\frac{\Delta}{2}, \Delta-2 ; \frac{\kappa}{2}\left(z^{2}+|\vec{x}|^{2}\right)\right) \label{e61}
\end{equation}

The imperfections of AdS/CFT applications in various condensed matter and quantum many-body theories often stems from the fact that the boundary CFT is a conformal field theory—even soft-wall holographic AdS/CFT remains a relatively idealized and simplified version. However, when the boundary CFT is coupled to matter in the AdS bulk via the boundary-to-bulk propagator and the dilaton, such modifications that alter form factors and coupling behaviors can instead yield an effective and realistic theory. This approach consistently couples quasiparticles or quasiparticle-like systems (including structures constructed from various phenomenological models in QCD) to background fields generated by gauge interactions and their corrections (provided by color kinematic duality or by correspondence).

\section{Exact Relations for $Z_{CFT}$ and $S_{AdS}$} \label{sec4}

\subsection{Strength Operator on Fiber Bundles}

In soft wall AdS/CFT, boundary CFT can be modified to obtain an effective (to annex loop contribution) running coupling via dilaton, and obtain effective quasiparticle-like phenomenological models after its form factor propagated into AdS bulk, thus physical system in AdS bulk has two equivalent description as discussed in part III. The source J in CFT mapped to field in AdS, while correlator O propagated into AdS bulk via boundary-to-bulk propagator, which gives a double-copy-like expression. And if we factorize physical system describing matter in AdS bulk, the form factor of effective current in AdS bulk is determined by boundary-to-bulk propagator, thus two sides of double-copy-like description $J_C \otimes \mathcal{O} \otimes G_{\partial-b} (x-y) J_{\text{eff}} (y)$ exchanges side on the opposite side of AdS/CFT mapping $J_{\text{bulk}} \otimes K_{b-b} (x-y) A_{\text{bulk}} (y), A_{\text{bulk}} \sim J_C \otimes G_{\partial-b}$.

Thus according to (whatever which side of) this relation we can convert strong currents or fields to weak currents or fields (or vise versa) just like AdS/CFT does, while exchange the role of two sides of double-copy-like description, and particle description and quasiparticle-like description should also exchanged. Define a "transformation operator" converting a field to another with different coupling while form factor also modified, which is a strength transforming operator

\begin{equation}
\mathcal{S}|_{g_0 \rightarrow g_1} : \sum_{L=0}^L A_{\mu, L} (x, g_0) e^{-L \phi(z_0)}=\sum_{L=0}^L A_{\mu, L} (x, g_1) e^{-L \phi(z_1)} \frac{ \Phi(x,y) \star \tilde{A_{1 \nu}}(x, \frac{a(\Lambda_{\text{eff}})}{g_1}) }{\Phi(x,y) \star \tilde{A_{0 \nu}}(x, \frac{a(\Lambda_{\text{eff}})}{g_0}) } \label{e62}
\end{equation}

where $\tilde{A_{0 \nu}}$ and $\tilde{A_{1 \nu}}$ are determined by their form factor distribution

\begin{equation}
F(\tilde{A_{0 \nu}})=F(A_{\mu}(g_0)) e^{\phi(z_0) r_0/2 } U(a_0,b_0,z_0), F(\tilde{A_{1 \nu}})=F(A_{\mu}(g_1))e^{\phi(z_1) r_1/2 } U(a_1,b_1,z_1) \label{e63}
\end{equation}

The expansion of $\star$ product given by part II. $\phi(z)$ is dilaton, which is usually written as $\Phi(z) = \kappa^2 z^2$, but dilaton can actually be regarded as a modification, with its expression only needs to satisfy conditions \cite{ac2} such as approximate conformal invariance and defining a running coupling as mentioned, and produce linear Regge trajectories, among others, without violating the constraints of mathematical and physical principles.. And in various physical systems, ranging from QCD nuclear matter with self-interactions or confinement potentials to QED-based quantum many-body theories, the confluent hypergeometric function U(a, b, z) of the second kind is used here to describe the interaction of an effective single particle within a parameterized many-body background. By superimposing or integrating these descriptions according to the distribution statistics of such particles, one can characterize the overall spatial distribution of the physical system. $\mathcal{O}$ and $\mathcal{O}^*$ is given in part.II.

In fiber bundle we can describe this relationship with $\omega = i g A^a_\mu dx^\mu T_a$, $ \Omega = d\omega + \omega \wedge \omega$, and with a pullback we have

\begin{equation}
A = A_{\mu} dx^{\mu} = A_{\mu}^a T^a dx^{\mu}, F = dA + A \wedge A,  S_{YM} = -\frac{1}{2g^2} \int_M \mathrm{Tr} (F \wedge \star F) \label{e64}
\end{equation}

Where $\wedge \star$ is hodge dual. As $(G, g_0, \mu) \rightarrow (G, g_1, \mu)$, section of associated bundle should transform as

\begin{equation}
\omega_0 = i g_0 A^a_\mu dx^\mu T_a \rightarrow \omega_1 = i g_1 A^a_\mu dx^\mu T_a, \text{ while } \psi_0 = e^{\frac{1}{2}\phi(g_0, g_1, \mu)} U(a,b,z) \psi_1 \label{e65}
\end{equation}

To represent the same physical system. This is a preparation for following discussion.

\subsection{Form Operator with Bundles Description}

Physical system usually contains not only one types of gauge field and interact, but also affected by various external fields, and kinematic factors, which is a reason making quantum many-body system complex, and a reason making non-pertubative QCD difficult as well. In revised fiber bundle it combines various contribution on 4-form $\text{Tr}(F \wedge *F) = \mathcal{L} dV$, and $Z = \int \mathcal{D}\phi e^{-\int dV \mathcal{L}}$.

In the AdS/CFT correspondence, the global symmetries of various boundary CFT correspond to gauge symmetries in the bulk. To ensure the equivalence, the symmetries should transform without loss of information. Thus a compact microscopic-dimensional Kähler manifold is introduced within the AdS bulk. In the double copy, similar mechanics is described by color-kinematic duality.

With Kähler manifold whose metric is a Kähler metric $K_{i\bar{j}} = \partial_i \partial_{\bar{j}} K$ on its complex coordinates $z^i, i = 1, 2, \ldots, n$, we locally define a Kähler potential $K(z, \bar{z})$ and Kähler form $\omega = i K_{i\bar{j}} dz^i \wedge d\bar{z}^{\bar{j}}, d\omega = 0$. Rewrite kinetic lagrangian

\begin{equation}
\mathcal{L}_{\text{kin}} = M_P^2 \frac{\partial^2 \hat{K}}{\partial \phi^i \partial \bar{\phi}^{\bar{j}}} D_\mu \phi^i D^\mu \bar{\phi}^{\bar{j}}, \text{dimensionless Kähler potential} \hat{K} = \frac{K}{M_P^2} \label{e66}
\end{equation}

Where $K_{i\bar{j}} = \frac{\partial^2 K}{\partial \phi^i \partial \bar{\phi}^{\bar{j}}}$ is Kähler metric, and $M_P$ is reduced Planck Mass. Also $K_{i\bar{j}} = \langle O_i O_{\bar{j}} \rangle$, $O$ is correlator, and pure kinematic contribution on gravitational system enacted by $O$ and $K_{i\bar{j}}$. With the compact microscopic-dimensional Kähler manifold, we can trigger various gauge interacts via color-kinematic duality.

Previous works \cite{2309.03289} added a "weak constraint" on the double copy by adopting double field theory. For spin-s level, We generate $A(J)$ from kinematic background via

\begin{equation}
\langle J_s(x_a) J_s(x_b) \rangle = C_s P^{(s)}_a P^{(s)}_b H^s_{ab} \langle O(x_a) O(x_b) \rangle \label{e67}
\end{equation}

Where weight-shifting and spin-raising operators \cite{2104.12803} $H_{ab} = 2(\vec{z}_a \cdot \vec{K}_{ab})(\vec{z}_b \cdot \vec{K}_{ab}) - (\vec{z}_a \cdot \vec{z}_b) K^2_{ab}$, K is weight-shifting operator, P is spin-s projector based on Todorov Operator \cite{2104.12803}. Kinematic corrections are classified as an effective anomalous "magnetic moment" to give external field correction $\Delta \mu_{\text{Kähler}} = \lambda  g_{\text{eff}}(\frac{1}{\mathcal{V}})  K_{i\bar{j}}  \langle O^i O^{\bar{j}} \rangle_{p=0}$ on scattering amplitudes,

$\mathcal{V}$ is volume of Kähler manifold. While for the field-strength distribution ($J_\mu^a A^{a\mu}$ part is discussed above),

\begin{equation}
\mathcal{L}_{\text{kin}} = K_{i\bar{j}} \partial_\mu \phi^i \partial^\mu \bar{\phi}^{\bar{j}} \rightarrow \mathcal{L}_{\text{eff}} = -\frac{1}{4} \text{Im}(\mathcal{N}_{AB}) F_{\mu\nu}^A F^{\mu\nu B} \label{e68}
\end{equation}

Where $\mathcal{N}_{AB}$ is a matrix determined by the dual Kähler potential. And in AdS bulk the variable $x$ is actually $(x, z)$, the modifier of distribution of form factors is along with the change of scale, $F_1 (x,z) = \left[ e^{\phi(z)} \right]^{-d} F_0\left( \frac{x}{e^{\phi(z)}} \right)$, z in AdS bulk related to energy scale $\Lambda$. and

\begin{equation}
\begin{aligned}
\mathcal{F}_{J}|_{0 \rightarrow 1} \langle J_s(x_a) J_s(x_b) \rangle_{\mu,0} &= \langle J_s(x_a) J_s(x_b) \rangle_{\mu,1} \frac{C_{s1} P^{(s)}_{a1} P^{(s)}_{b1} H^s_{ab1}}{C_{s0} P^{(s)}_{a0} P^{(s)}_{b0} H^s_{ab0}}\\
\mathcal{F}_{O}|_{0 \rightarrow 1} F(R_0, \frac{x}{e^{\phi(z_0)}}) \wedge *F(R_0, \frac{x}{e^{\phi(z_0)}}) &=F(R_1, \frac{x}{e^{\phi(z_1)}}) \wedge *F(R_1, \frac{x}{e^{\phi(z_1)}})
\label{e69}
\end{aligned}
\end{equation}

$R_0$ is scale of Kähler manifold $\text{Ric}_{i\bar{j}} = -\partial_i \partial_{\bar{j}} \log \det(g_{k\bar{l}})$. Summarize systems across various fields via form operators

\begin{equation}
\begin{aligned}
\sum \mathcal{F}|_{F_n \rightarrow F_0} A_{ n}\left(\langle JJ \rangle_{ n}, O(F \wedge *F)_n \right)=&A_{ 0}\left(\sum \mathcal{F}_{J}|_{n \rightarrow 0} \langle JJ \rangle_{ n}, \sum \mathcal{F}_{O}|_{n \rightarrow 0} O(F \wedge *F)_n \right)\\
=&A_{ 0}\left(\langle JJ \rangle_{0, \text{eff}}, O(F \wedge *F)_{0, \text{eff}}\right)
\label{e70}
\end{aligned}
\end{equation}

This is a degenerate version of T-duality based on color-kinematic duality, while changing spatial scales due to the compactification radius, and results in changes to the energy scale. Add this duality on the bundle structure described in previous subsection, we can combine principal bundles with different symmetry groups and their associated bundles into a united structure if we have a Product Operator.

\subsection{Product Operator with Kernel Defined by Kähler Manifold}

In the double copy, we need a compact microscopic-dimensiona Kähler Manifold to describe the symmetry of the kernel in gravitational system. This manifold is jointly determined by $A_{\mu, \text{eff}}^{\text{no-loop}}(x)$ side and $\tilde{A}_{\nu , \mathcal{O}}^{\text{form invariance}}(x)$ side in complete system with gravity. This manifold evolves effectively degenerates to a point at $A_{\mu, \text{eff}}^{\text{no-loop}}(x)$, as it does in AdS/CFT as $z \rightarrow 0$. In soft wall AdS/CFT, as we introduce the double copy with specific color-kinematics duality during the scale-operators (operators refers to $J$ and $O$) evolving, effective geometry of "effective" Kähler manifold X (applicated in form operator $\mathcal{F}$ in subsection above) for probes of the boundary CFT becomes

\begin{equation}
K_{\text{eff}}(z) = e^{\alpha \Phi(x,z)} \frac{R^2}{z^2} f(z) K_X + \text{quantum corrections}( \sim \frac{1}{c} \Phi(x,z) K_X + \mathcal{O}(\Phi^2)) \label{e71}
\end{equation}

c is the CFT central charge, and dilaton given by $\Phi(x,z)$. Kähler manifold describes the kernel between various gauge-field-based bundles, modifies the form factors in $A_{\mu, \text{eff}}^{\text{no-loop}}(x)$ as

\begin{equation}
F(q^2, z) = F_0(q^2) \cdot \exp\left[\mathcal{K}_{\Phi}(q^2, z) \cdot \mathcal{G}_{\Phi}(g_{i\bar{j}}, \omega, K, \Phi) \cdot \mathcal{H}(\Phi)\right] \label{e72}
\end{equation}

Where $\mathcal{K}_{\Phi}(q^2, z)$ encodes the momentum-dependent contributions of KK modes to the form factor, $\mathcal{G}_{\Phi}(g_{i\bar{j}}, \omega, K, \Phi)$ encodes how the boundary global symmetry emerges from the bulk gauge symmetry and how the Kähler geometry modifies this correspondence, $\mathcal{H}(\Phi)$ is additional dilaton form factor.

The specific double copy in present work, is actually factorizing a physical system into a product (with convolution) form of one effective field annexed all loop or self-interact contribution in effect couplings $A_{a, \mu}(J_{\mu},O)$, and another effective field using emerging effective particles describing their loop or self-interact contribution (but is described locally only in region inside this effective model) $A_{b, \nu}(J,O_{\nu})$, combined via a product operator

\begin{equation}
g_{\mu \nu}=A_{\mu}(J_{\mu},O) \mathcal{P}\left\langle A_{a, \mu}, A_{b, \nu} \right\rangle A_{\nu}(J,O_{\nu})
\label{e73}
\end{equation}

And when we refocus on phenomenological approaches, the phenomenological version of unified duality ensures the model-independence with two types of corrections: Corrections reflected in the effective selection of couplings for the gravitational system, gauge couplings, and mass term couplings (Yukawa coupling); corrections reflected in the selection of the effective model and its symmetries. Thus we give formula of product operator

\begin{equation}
\begin{aligned}
\mathcal{P}\left\langle A_{a, \mu}, A_{b, \nu} \right\rangle &= \star \Phi_0 \mathcal{K}_{\text{coupling}} \mathcal{K}_{\text{model}} \cdot \left( \det K_{i\bar{j}} \right)^{-1/2} \cdot \exp\left( -\frac{1}{2} K(z, \bar{z}) \right) \star \\
\mathcal{K}_{\text{coupling}}&=\exp\left[ -\mathcal{V}_M \int_M \left( \omega_{K} \wedge \omega_{K} \wedge \omega_{K} + \alpha_1 R \wedge R + \alpha_2 \text{tr}(F \wedge F) \right) \right]\\
\mathcal{K}_{\text{model}}&=\chi(M) \cdot |\text{Aut}(M)| \cdot \exp\left[ 2\pi i \int_M c_1(M) \wedge \omega_{K} \right]
\label{e74}
\end{aligned}
\end{equation}

F refers to the field strength F of the macroscopic physical system, $\mathcal{V}_M$ is the volume $\mathcal{V}_M = \frac{1}{3!} \int_M \omega_{K} \wedge \omega_{K} \wedge \omega_{K}$,and R is curvature form of Kähler Manifold, and $\omega_{K}$ here refers to Kähler form $\omega_{K} = i K_{i\bar{j}} dz^i \wedge d\bar{z}^{\bar{j}}$. This result describe effects from $g_{\text{eff}}$ and $\Lambda_{\text{eff}}(x, \mathcal{K})$ on kernel, and 

\begin{equation}
\Lambda_{\text{eff}}(x, \mathcal{K})= \Lambda_0 \left[ 1 + \sum_{k=1}^{L} \left( c_k^{(T)} \left( \frac{T}{\Lambda_0} \right)^{2k} + c_k^{(\mu)} \left( \frac{\mu}{\Lambda_0} \right)^{2k} + \frac{c_k^{(\mathcal{L})}}{\Lambda_0^{4k}} \mathcal{L}_{\text{ext}}^k \right) + O(\text{higher}) \right] \label{e75}
\end{equation}

Where $\mathcal{L}_{\text{ext}} = J \cdot O$ is directly defined by $\mathcal{K}$ as discussed above. Especially, we simplify it for $A_{\mu}(J_{\mu},O)$

\begin{equation}
\Lambda_{\text{eff}}(T, \mu, \mathcal{L}_{\text{ext}}) = \kappa \sqrt{ 1 + b \left( \frac{T}{\kappa} \right)^2 + d \left( \frac{\mu}{\kappa} \right)^2 + h \frac{\mathcal{L}_{\text{ext}}}{\kappa^4} } \label{e76}
\end{equation}

In this revised bundle structure, the product operator is defined to induce the tensor field $g_{\mu \nu} (x)$ on base manifold(from fields on bundles), and give Kähler-Manifold-dependent kernel to unify bundles. Levi-Civita connection on base manifold

\begin{equation}
\Gamma^\lambda_{\mu\nu} = \frac{1}{2} g^{\lambda\rho} \left( \partial_\mu g_{\nu\rho} + \partial_\nu g_{\mu\rho} - \partial_\rho g_{\mu\nu} \right), R_{\mu\nu} - \frac{1}{2} R g_{\mu\nu} + \Lambda g_{\mu\nu} = \frac{8\pi G}{c^4} T_{\mu\nu}
\label{e77}
\end{equation}

Unify the double copy perspective and AdS/CFT perspective, the revised bundle structure utilizes operators defined above to arrive at an expression, that aggregates the contributions from various effective gauge fields, a commutative ring emerges: Operator $\mathcal{F}$ adjust n-forms of various type of field that they can do "add" operation, and operator $\mathcal{P}$ defines the product, and influence the base manifold by inducing $\Gamma^\lambda_{\mu\nu}$ and $g_{\mu\nu}$, create a commutative ring connection bundle.

\subsection{Commutative Ring Perspective and Phenomenological Result}

To give phenomenological predicts, we derive Hamiltonian density from the bundle structure via Legendre transformation (as $^*$)

\begin{equation}
\begin{aligned}
\mathcal{H}(x) &= \left[ - \text{Tr} (F \wedge *F) \right]^* + O(x) - J^i(x) A_i(x)\\
&=\text{Tr} \left[ F_{0i}(x) F_{0i}(x) \right] + \frac{1}{2} \text{Tr} \left[ F_{ij}(x) F_{ij}(x) \right] + O(x) -  J^i(x) A_i(x)
\label{e78}
\end{aligned}
\end{equation}

A is 1-form, F is 2-form, O and J are operators discussed above. And for Hamiltonian contribution from gravitational system

\begin{equation}
H_{\text{total}} = \int_\Sigma d^3x (N\mathcal{H}_0 + N^i\mathcal{H}_i) + H_{\text{boundary}}, H_{\text{boundary}} = -2 \int_\infty dS_i (\partial_j h_{ij} - \partial_i h)
\label{e79}
\end{equation}

Where N is Lapse Function, $\mathcal{H}_0$ is Hamiltonian Constraint, $N^i$ is Shift Vector, and $\mathcal{H}_i$ is Momentum Constraint.

And ADM mass $E_{ADM}=\frac{1}{16 \pi} \lim_{r \rightarrow \infty} \int_S \left( \partial_j h_{ij} - \partial_i h_{jj} \right) N^i dS= \frac{-1}{8\pi} H_{\text{boundary}}$ in asymptotically flat spacetime as $N \rightarrow 1$ and $N^i \rightarrow 0$.

In this subsection, Hamiltonian is defined on bundle structure, connects various physical systems by correspondences. Previous researches have discussed gravity with loops \cite{2110.14688}\cite{9504092}, they are now given via that specific form of double copy. In this perspective, $A$ and $J$ evolves as 1-form-level $\omega$, $F_{\mu \nu}F^{\mu \nu}$ and $O$ after the order of $F \wedge *F$ are 4-form-level integrands and actually evolves as $F \wedge *F d^3 x$, and $ (\omega_{A,\mu} \omega_J^\mu) (- d^4x)=\omega_A \wedge \star \omega_J $. Their transformation rules on the according to the commutative ring which is defined in the commutative ring connection bundle:

\begin{equation}
\begin{aligned}
\mathcal{P}\left\langle (\omega_0,F_0), (\omega_1,F_1) \right\rangle &= \star \Phi_0 \mathcal{K}_{\text{coupling}} \mathcal{K}_{\text{model}} \cdot \left( \det K_{i\bar{j}} \right)^{-1/2} \cdot \exp\left( -\frac{1}{2} K(z, \bar{z}) \right) \star\\
\mathcal{F}|_{O \rightarrow J}: K_{i\bar{j}} \partial_\mu \phi^i \partial^\mu \bar{\phi}^{\bar{j}}  &\sim F_0 \wedge *F_0 dV \left(F_0 \wedge *F_0 dV \right), \quad C_{s} P^{(s)}_{a} P^{(s)}_{b} H^s_{ab} \rightarrow \omega_J \cdot \omega_A  \\
\mathcal{S}_{\omega, F}|_{\text{Standard} \rightarrow \text{Effective}}:& \underbrace{\left(\omega_A \wedge \star \omega_J+F \wedge *F d V\right)_{g_0}}_{S_0} = \underbrace{\left[\omega_A \wedge \star \omega_J \left( \frac{c F(x)}{r_{\text{eff}}} e^{m_{\text{eff}}} U(a, b, z)\right)\right]_{g_1}}_{S_1}
\label{e80}
\end{aligned}
\end{equation}

\begin{equation}
g_{\mu \nu} (x)=  \sum_{L=0}^L S_{0, \mu, L} e^{-L \phi(z)} \mathcal{P}\left\langle (\omega_0,F_0), (\omega_1,F_1) \right\rangle \sum_{L=0}^L \tilde S_{1, \nu, L} e^{L \phi^*} \label{e81}
\end{equation}

Here $*$ is adjoint operator defined by Killing form. Utilizing Hamiltonian evolves under $\mathcal{P}\left\langle (\omega_0,F_0), (\omega_1,F_1) \right\rangle, \mathcal{F}|_{O \rightarrow J}$, and $\mathcal{S}_{\omega, F}$, we have

\begin{equation}
\begin{aligned}
Z[\eta] = \int \mathcal{D} A_i \exp \left( - \int d^4x_E \right. &\left. \left[ \text{Tr}  \left[ (\partial_4 A_i)^2 \right] + \frac{1}{2} \text{Tr} \left[ (\partial_i A_j - \partial_j A_i - i [A_i, A_j])^2 \right] + O(x) \right. \right.\\
\left.- J^i(x) A_i(x) \right]&\left.+ \int d^4x_E \eta^i(x) A_i(x) \right) \text{(real-time)}\\
Z_{\tau} = \int \mathcal{D}A_i \mathcal{D}\phi  \exp \left[ - \int_0^\beta d\tau \right. &\left. \int d^3x \left( \frac{1}{2} \text{Tr} [ (\partial_\tau A_i)^2 ] + \frac{1}{2} \text{Tr} [ F_{ij} F_{ij} ] + L_{\text{matter}} - J^i A_i \right) \right] \\
\langle A_{i1}(x1) \cdots A_{in}(xn) \rangle &= \frac{1}{Z[0,0]} \frac{1}{i} \frac{\delta}{\delta \alpha_{i1}(x1)} \cdots \frac{1}{i} \frac{\delta}{\delta \alpha_{in}(xn)} Z[\alpha,0] \bigg|_{\alpha=0}\\
\langle \phi(y1) \cdots \phi(ym) \rangle &= \frac{1}{Z[0,0]} \frac{1}{i} \frac{\delta}{\delta \eta(y1)} \cdots \frac{1}{i} \frac{\delta}{\delta \eta(ym)} Z[0,\eta] \bigg|_{\eta=0}\\
\langle J^i(x) J^j(y) \rangle &= \frac{1}{Z} \int \mathcal{D}A_i \mathcal{D}\phi \, J^i(x) J^j(y) \exp[i S] \quad \text{(real time)}\\
\Omega = - T \ln Z, S = - \frac{\partial \Omega}{\partial T}, T_{ij}(x) &= \text{Tr} [ F_{i\mu} F_{j\mu} ] - \frac{1}{4} \delta_{ij} \text{Tr} [ F_{\mu\nu} F_{\mu\nu} ] + T_{ij}^{\text{matter}}\\
\label{e82}
\end{aligned}
\end{equation}

Where $\eta^i(x)$ is an auxiliary source introduced to compute correlation functions, sharing the same dimension and similar physical role as $J^i(x)$, and $Z_{\tau}$ is imaginary-time generating function (partition function).

\section{Conclusions} \label{sec5}

In quantum many-body system of QCD, the strong coupling constant and the color confinement restrict the range of interactions and cause some simplicity, while this confinement simultaneously imposes an intricate internal structure that demands rigorous analyzing. Conversely, in QED the coupling constant assumes a modest value, while almost infinite number of interacting entities causes complexity for analyzing various states of matter it affects.

In this work, we established correspondence relations between equivalent descriptions connecting particles and quasi-particles or other effective models, bridging choices of sources from various symmetries, and spanning boundaries of distinct background fields and ranges of effective models, while preserving model independence, with background spacetime effects under consideration. Thus we can circumvent the redundant degrees of freedom and complexities and circumvent difficulties in calculating them. Further more, it enable more effective and accurate predictions of physical system states and their thermodynamic evolution based on a complete theoretical foundation.

\normalem


\end{document}